\documentclass{Rinton-P9x6}
\def\f{\frac}
\def\bea{\begin{eqnarray}}
\def\eea{\end{eqnarray}}
\def\nn{\nonumber}
\def\la{\langle}
\def\ra{\rangle}
\newcommand{\GeV}{\,{\rm GeV}}
\begin{document}

\title{Gauge Coupling Unification and Neutrino Masses in 5D SUSY SO(10)}

\author{Stuart Raby\footnote{This talk is based on work in collaboration
with H.D. Kim}}

\address{Department of Physics, The Ohio State University,\\ 174 W.
18th Ave., Columbus, OH 43210, USA
\\E-mail: raby@pacific.mps.ohio-state.edu
}

\maketitle

\abstracts{ In this talk I discuss the problems and virtues of
SUSY GUTs in four dimensions.  I then show how to solve some of
these problems,  without foregoing the virtues,  by considering an
SO(10) SUSY GUT in five dimensions.   I discuss gauge coupling
unification and neutrino masses.   In particular, it is shown that
the 5D compactification scale and cutoff scale are determined by
fitting the low energy values of the three standard model gauge
couplings.  Neutrino masses can be accommodated.}

\section{Gauge coupling unification in 5D}

We consider gauge coupling unification in $SO(10)$ in five
dimensions~\cite{Kim:2002im}.   In particular we discuss hybrid
gauge symmetry breaking with both orbifold and Higgs vevs on the
brane. We calculate the GUT scale threshold corrections to gauge
coupling unification.  We then show that the compactification
scale $M_c \approx 10^{14} \; {\rm GeV}$ and the cutoff scale $M_*
\approx 10^{17} \; {\rm GeV}$ are fixed by the low energy data.
Finally we consider neutrino masses and determine the  See -- Saw
scale determining light neutrino masses~\cite{Kim:2003vr}.  Let us
first define some notation.

\subsection{Charge quantization \& Family structure}\label{subsec:chquant}

The Pati--Salam gauge symmetry $SU(4)_c \times SU(2)_L \times
SU(2)_R$ unifies quarks and leptons of one family into two
irreducible representations given by
$$ \bf \psi \;\; = \;\; (4, 2, 1) \;\; = \;\; \{  Q = \left(\begin{array}{c}
u \\  d \end{array} \right), \;\;\;  L = \left(\begin{array}{c}
\nu \\  e
\end{array}\right) \} $$
and
$$ \bf \psi^c \;\; = \;\; (\bar
4, 1, \bar 2) \;\; =  \;\; \{   Q^c = \left(\begin{array}{c} u^c
\\  d^c
\end{array} \right), \;\;\;
  L^c = \left(\begin{array}{c} \nu^c \\ e^c \end{array}\right)  \} . $$
The two Higgs doublets of the minimal supersymmetric standard
model are contained in one irreducible representation
$$\hspace{.5in}   {\cal H} \;\; =  \;\; (1, \bar 2, 2) \;\; =\;\;  \{   H_u ,\;\; H_d \} .$$
Hence Pati-Salam naturally describes the family structure of the
standard model.  Moreover since there are no U(1) symmetries,
charge quantization is enforced. There are however three
independent gauge couplings [two if one also demands parity] and
thus no prediction for gauge coupling unification.

The gauge group SO(10) then unifies quarks and leptons into one
irreducible spinor representation
$$\hspace{.5in}  \bf \psi \;\; + \;\; \bf \psi^c \;\;\; \subset \;\; {\bf 16}. $$
With the addition of Higgs triplets, the Higgs doublets are
contained in the defining representation
$$\hspace{.5in}  {\bf {\cal H}} \;\; + \;\;  (6, 1,
1) \ ( = \{  T, \;\; \bar T \} )
  \;\; \subset \;\; {\bf 10_H} .$$   Of course, SO(10) also
predicts gauge coupling unification.

Finally both symmetry groups, Pati-Salam and SO(10), lead
naturally to Yukawa unification for the third generation with
$$\hspace{.5in} \lambda \ 16_3 \ 10_H \ 16_3 \ \supset \ \lambda \ \bf
  \psi_3 \; {\cal H} \; \psi_3^c $$ and a single Yukawa coupling
  $\lambda$.   Given the above brief review, let us consider the
  virtues and problems of four dimensional SUSY GUTs.

\subsection{Virtues and problems of 4D SUSY GUTs}

Four dimensional SUSY GUTs have the following virtues.
\begin{itemize}
\item Charge quantization --- {\it No U(1) factors}

\item Family structure --- {\it Quarks and leptons are in the
smallest chiral (i.e. non vector-like) representations}

\item  Neutrino Mass [ $\nu \ m \ \nu^c \ + \ \frac{1}{2} \ \nu^c
\ M \  \nu^c $] with  $m = \lambda \langle H_u \rangle$ and  $M$ [
=  See--Saw scale] $\sim M_{GUT}$, we obtain $\; m_\nu \; = \;
m^2/M$ --- {\it A right-handed neutrino $\nu_R = (\nu^c)^*$ is
required in either PS or SO(10)}

\item  Gauge coupling unification ---  {\it Fits the low energy
data}

\item  Yukawa coupling unification --- {\it This is a prediction
of minimal PS and SO(10)}

\item  Dark matter candidate  --- {\it With a conserved R-parity,
the LSP (typically the lightest neutralino ($\tilde \chi^0_1$)) is
stable}

\end{itemize}

\bigskip

4D SUSY GUTs have the following problems.

\begin{itemize}
\item  Gauge symmetry breaking requires a complicated symmetry
breaking sector.

\item  Higgs doublet -- triplet splitting can be accommodated but
is not required by the theory.

\item  A supersymmetric $\mu$ term, with a dimensionful parameter
of order the electroweak scale, must be generated.

\item  Proton decay, due to dimension 5 operators, must be
suppressed to satisfy the Super-Kamiokande bound - 1/$\Gamma(p
\rightarrow K^+ \ \bar \nu) > 1.9 \times 10^{32} \; {\rm yrs}$.

\item  In order to obtain Majorana neutrino masses consistent with
atmospheric neutrino oscillations with $\Delta m^2_{atm} \sim 3
\times 10^{-3} \; {\rm eV}^2$, one needs a See-Saw scale $M \ \sim
\ 10^{-2} \ M_{GUT} \ll M_{GUT}$ for the tau neutrino (assuming
the light neutrino spectrum is hierarchical).
\end{itemize}  We now consider SUSY SO(10) on an orbifold in 5D and show how some
of these problems can be resolved, while at the same time
retaining the virtues of 4D SUSY GUTs.

\subsection{SUSY SO(10) on ${\cal M}_4 \ \times \ S_1/(Z_2 \times
Z_2^\prime)$}

The 5D orbifold is a line segment [0, $\pi R/2$] in the fifth
dimension $y$ defined in terms of a $Z_2 \times Z_2^\prime$
orbifolding of the circle with radius $R$.  The first $Z_2$ breaks
the effective 4D N=2 SUSY to N=1 SUSY, while the second breaks
SO(10) to Pati-Salam [PS].\footnote{Orbifold breaking of SO(10) in
5D was discussed in Ref.~\cite{Dermisek:2001hp} or in 6D in
Refs.~\cite{Asaka:2001eh}.  We see no advantage in going to 6D.}
Hence the brane at $y = 0$ has the full SO(10) symmetry, while the
brane at $y = \pi R/2$ has only the PS symmetry.  The 5D bulk
fields (given in terms of 4D superfields) include the gauge sector
[$V, \Phi$] in the adjoint representation and the Higgs
hypermultiplet [$10_H, 10_H^c$] in the defining representation. In
a standard notation we then have the $Z_2 \times Z_2^\prime$
eigenstates $V_{+ +}, \ \Phi_{- -} \;\; \subset$ PS;  $\;\; V_{+
-}, \ \Phi_{- +} \;\; \subset$ SO(10)$/$PS; $\;\; {\cal H}_{+ +},
\ {\cal H}^c_{- -} \;\; \subset \;\; (1,\bar 2, 2) \;$ and $ \;
T_{+ -}, \ \bar T_{+ -}, \  T^c_{- +}, \ \bar T^c_{- +} \;\;
\subset \;\; (6, 1, 1)$. {\em Thus only $V_{+ +}$ (the PS gauge
sector) and ${\cal H}_{+ +}$ (the Higgs doublets) contain zero
modes.}   We then assume that the fields $\langle \chi^c \rangle =
\langle \bar \chi^c \rangle$, located on the PS brane, develop a
vev of order $ \sim M_*$ [
 = cutoff scale];  spontaneously breaking PS to the
standard model gauge group.  The three families of quarks and
leptons live either on the PS or SO(10) brane or in the bulk. They
come in complete families either under SO(10) or PS.   With this
construction our 5D theory has the properties described in the
following theoretical score card. [$\surd$, means it is a property
of the construction, while {\bf ?}, will be discussed further in
this talk.]

\subsection{5D SO(10) --- Theoretical Score Card}

\begin{itemize}
\item Charge quantization \& Family structure --- $\surd$
\medskip

\item Gauge coupling unification --- {\bf ?}
\medskip

\item Yukawa coupling unification for the third generation ---
$\surd$
\medskip

\item R parity $\Longrightarrow$ dark matter candidate --- $\surd$
\medskip

\item Neutrino mass (See--Saw mechanism) ---  {\bf ?}
\medskip

\item Gauge symmetry breaking --- $\surd$
\medskip

\item Higgs doublet--triplet splitting --- $\surd$
\medskip

\item Proton decay ($p \rightarrow K^+ \ \bar \nu$) due to dim. 5
Operators --- {\it R symmetry prevents dim. 5 ops.}
--- $\surd$ \smallskip

\item Proton decay ($p \rightarrow e^+ \ \pi^0$) due to dim. 6
operators --- {\it negligible in 4D, however in 5D one is now
sensitive to physics at the cutoff and the effects are
incalculable (and perhaps even observable ?)}
\medskip

\item Right-handed neutrino mass scale ---  {\bf ?}
\end{itemize}

\subsection{Gauge coupling unification : Orbifold/Brane breaking}\label{sec:hybridbreaking}

We use orbifold breaking to explicitly break SO(10) to PS and
spontaneous breaking on the PS brane to break PS to the standard
model~\cite{Kim:2002im}.  We now calculate the threshold
corrections to gauge coupling unification, due to the tower of
Kaluza-Klein modes. We show that the cutoff scale, $M_*$,  the
compactification scale, $M_c$, and the value of the unified gauge
coupling at $M_*$ are fixed by the low energy data.   This differs
from 4D GUTs in only one respect.   There too a precise fit to the
low energy data requires three parameters, the GUT scale, $M_G$,
the gauge coupling at the GUT scale, $\tilde \alpha_G$, and a
perturbative threshold correction at $M_G$ due to the GUT and
Higgs breaking sectors of the theory.   If we define the GUT scale
as the point where $\alpha_1 = \alpha_2 = \tilde \alpha_G$, then
the necessary threshold correction is given by $\epsilon_3 =
(\alpha_3(M_G) - \tilde \alpha_G)/\tilde \alpha_G \sim - 4\%$.  In
the 5D case the ``GUT scale" threshold corrections are solely due
to the tower of KK states above $M_c$ and do not depend on
arbitrary parameters in the superpotential responsible for GUT
breaking.

Spontaneous symmetry breaking on the brane, i.e. brane breaking,
is accomplished with the vevs of two fields, $\chi^c$, and its
conjugate, $\; \bar \chi^c$, with the following transformation
under PS -- $\chi^c =  (\bar 4, 1, \bar 2)$.\footnote{For a nice
discussion of brane breaking, see Ref.~\cite{Nomura:2001mf}.}
Note, without loss of generality, $\chi^c$ is assumed to get a vev
in the right-handed neutrino direction $\chi^c \ \supset \ \nu^c $
which breaks PS to the standard model. This spontaneous breaking
generates a mass term for the PS gauge fields in PS/SM given by
$$\delta(y - \frac{\pi \ R}{2}) \;  g_5^2 \ (\langle \chi^c
\rangle^2 + \langle \bar \chi^c \rangle^2) \; A_\mu^2 .$$

\begin{figure}[t]
\epsfxsize=25pc 
\epsfbox{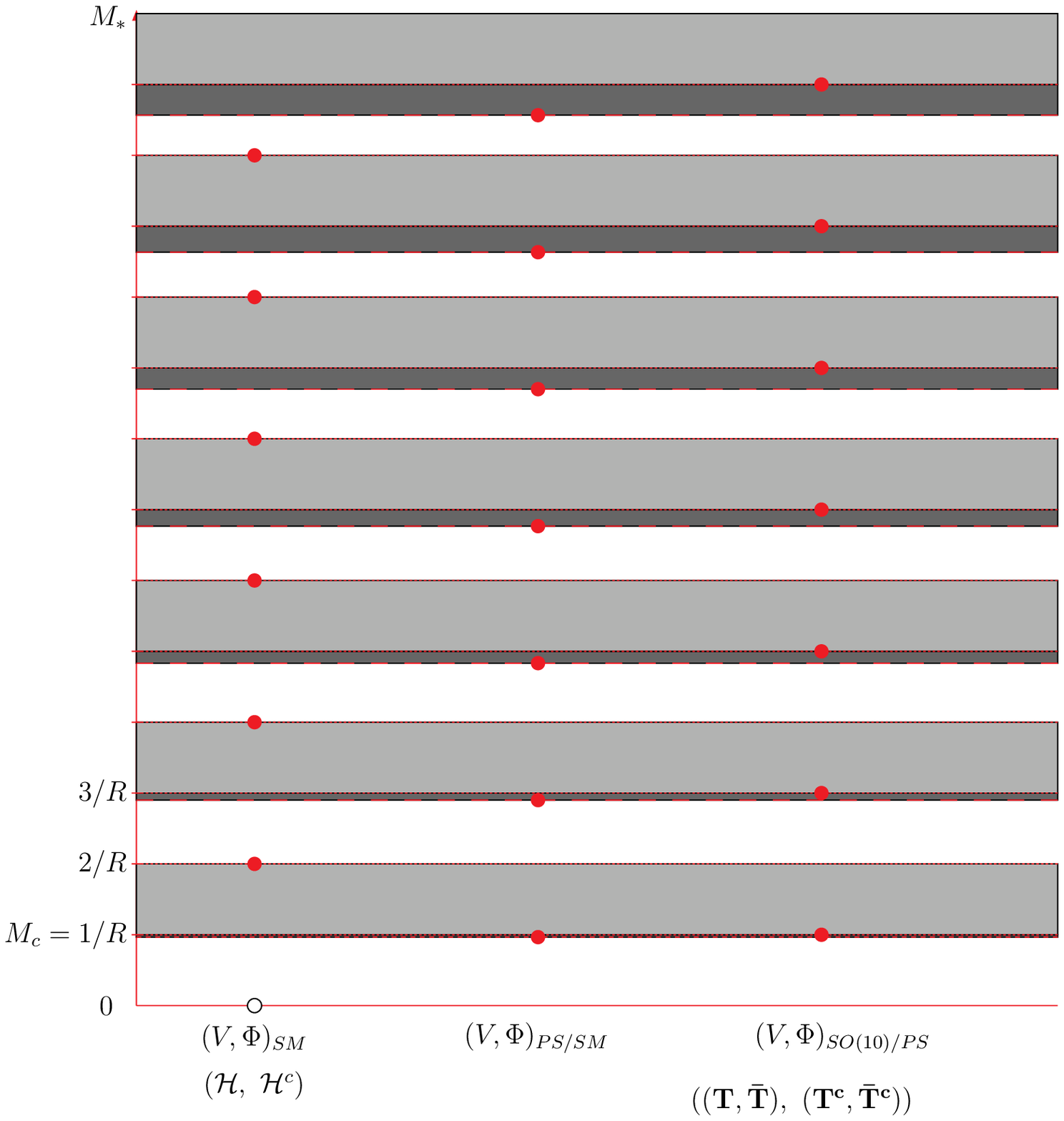} 
\caption{Gauge and Higgs spectrum. \label{fig:spectrum}}
\end{figure}

The brane mass terms affect the KK spectrum.   For $\langle \chi^c
\rangle \sim M_*$ the effect is to repel the wave functions of
fields in PS/SM away from the PS brane.  In particular, we find
$V_{+ +}, \ \Phi_{- -} \;\; \subset$ PS/SM goes to $\approx \;
V_{+ -}, \ \Phi_{- +}$, while  $V_{+ +}, \ \Phi_{- -} \;\;
\subset$ SM is unaffected.   The effect on $\Phi$ is a consequence
of the fact that N=1 supersymmetry is unbroken by the vevs.
Finally, the Higgs sector is unaffected by the brane breaking. The
resulting KK spectrum is illustrated in Fig. \ref{fig:spectrum}.

\subsection{$\Delta_i, \ i = 1,2,3$ --- Threshold corrections at $M_c$ due
to the KK tower between $M_c$ and $M_*$}

We are only interested in the running of the differences of gauge
couplings, hence we subtract an overall constant from $\Delta_i$
such that $\Delta_1 \equiv 0$ (for more details see
Ref.~\cite{Kim:2002im}).   The differences run logarithmically
above each KK mode.  The quarks and leptons come in complete
families and thus do not contribute at all to the running.  We
consider the gauge and Higgs contribution below.

The threshold corrections to gauge coupling unification defined at
the compactification scale are given by
 $  \Delta_i = \Delta_i^{gauge} \ + \ \Delta_i^{Higgs} $.
In the paper we have performed a detailed calculation (see also
Ref.~\cite{Hall:2001pg} for similar analyses). The exact result is
very well approximated by \bea \Delta_i^{gauge} \simeq &
\frac{2}{3} \ b_i^{SM}(V) \ \log (\frac{M_*}{M_c}) & \;\; {\rm and,} \\
\Delta_i^{Higgs} \simeq &  0 .& \eea This result is easy to
understand. Consider first the Higgs sector. When an equal number
of Higgs doublets and triplets contribute to the running, they
form a complete SO(10) multiplet and thus give zero contribution
to $\Delta_i$.  From Fig. \ref{fig:spectrum} we see that from
$M_c$ to $2 M_c$ there are more triplets than doublets. However
from $2 M_c$ to $3 M_c$ the situation is reversed with more
doublets than triplets.   These two cases alternate as one goes up
in energy and the net effect is to cancel each other. In the gauge
sector the situation is slightly different. Complete gauge or
$\Phi$ multiplets give zero contribution to the running. Again
from Fig. \ref{fig:spectrum} we see that from $M_c$ to $2 M_c$ the
$\Phi \subset$ SM contribution is missing, while from $2 M_c$ to
$3 M_c$ there is an excess of $V \subset$ SM. The net result gives
$2/3$ of the gauge sector contribution in the MSSM. With these
results we can now compare 5D and 4D SO(10).

\begin{figure}[t]
\epsfxsize=25pc 
\epsfbox{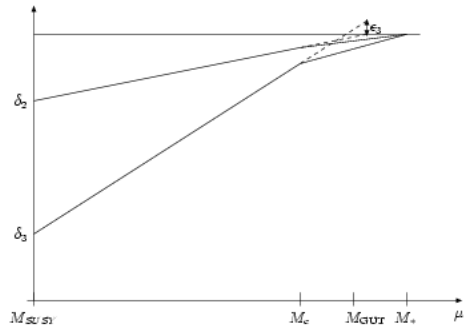} 
\caption{Differential running of $\delta_i = 2 \pi (
\f{1}{\alpha_i} - \f{1}{\alpha_1})$ for $ i = 2,3$.
\label{fig:running}}
\end{figure}

In 5D we have (for $\mu \leq M_c$) \bea \f{2\pi}{\alpha_i (\mu)} &
\simeq & \f{2\pi}{\alpha (M_*)} + b^{MSSM}_i \log ( \frac{M_c}{\mu}) +  \Delta_i \label{eq:5D} \\
& = & \f{2\pi}{\alpha (M_*)} + b^{MSSM}_i \log ( \frac{M_c}{\mu})
+ \f{2}{3} b^{SM}_i (V) \log (\f{M_*}{M_c}) \nn \eea   Whereas in
the 4D MSSM we have \bea \f{2\pi}{\alpha_i (\mu)} & \simeq &
\f{2\pi}{\alpha (M_{GUT})} + b^{MSSM}_i \log ( \frac{m_T}{\mu}) +
b^{SM}_i (V) \log (\f{M_{GUT}}{m_T}) \label{eq:4D} \eea Using two
loop RG running above the weak scale and one loop threshold
corrections at $M_Z$ we find $M_{GUT} \simeq 3 \times 10^{16} \;
{\rm GeV}$.  In addition the necessary GUT threshold correction,
$\epsilon_3$, can be obtained with the color triplet Higgs $T, \;
\bar T$ mass $m_T \simeq 2 \times 10^{14} \;\; {\rm GeV}$ (see for
example, Ref. ~\cite{Murayama:2001ur}).  From Eqns. (\ref{eq:5D},
\ref{eq:4D}) it is easy to see that the 5D RG equation is
equivalent to the 4D MSSM result if we take \bea M_c = m_T,
\,\,\,\,\, \f{M_*}{ M_c} = (\f{M_{GUT}}{m_T})^{\f{3}{2}},
\,\,\,\,\, \alpha(M_{GUT}) = \alpha(M_*) .
\label{eq:gcunification} \eea  We thus obtain gauge coupling
unification with $ M_c \simeq 10^{14} \; {\rm GeV}$ and $M_*
\simeq 10^{17} \; {\rm GeV}$.  The differential running of the
gauge couplings above $M_c$ is illustrated in Fig.
\ref{fig:running} for the 5D and 4D cases.

\section{Neutrino mass in 5D SO(10)}

For the See-Saw mechanism to work in a 4D SO(10) model, we need to
give the right-handed neutrino a Majorana mass of order the GUT
scale.   There are two methods for obtaining this:
\begin{itemize} \item using a higher dimension operator
\begin{equation} \f{\overline{16} \ 16 \ \overline{16} \  16}{M_*}, \;\;
 \end{equation} where $M_*$ is the cutoff scale of the theory, or
 \item adding an $SO(10)$ singlet $N$ and the
renormalizable interactions \begin{equation} \lambda_N \
\overline{16} \ 16 \ N + \f{1}{2} \ M_2 \ N \ N . \end{equation}
\end{itemize}
In both cases we assume a non-zero vacuum expectation value [vev]
$$\lambda_N \ \langle {\bf \overline{16}} \rangle  = M_1 \neq 0$$ in the right-handed neutrino
direction.   In the first case, the product of fields has, in
general, several inequivalent SO(10) invariant combinations. One
particularly simple combination in the first case, with the first
two fields combined to make an SO(10) singlet, can be obtained as
an effective interaction after integrating out $N$ in the second
case.

We now include the usual, electroweak scale, Dirac mass coming
from the Yukawa term $ \lambda \ 16_3 \ 10_H \ 16_3$ (with 3
denoting the third generation).  After electroweak symmetry
breaking we have \bea W & = & m_D \ \nu \ \nu^c \eea (with $m_D =
\lambda \f{v}{\sqrt{2}} \sin\beta$) and we obtain a $3 \times 3$
neutrino mass matrix, rather than a $2 \times 2$
matrix,\footnote{This is similar to the double see-saw mechanism
suggested by Mohapatra and Valle~\cite{doubleseesaw}.} given by
\bea  {\cal M} \left(%
\begin{array}{c}
  \nu \\
  \nu^c \\
  N \\
\end{array}%
\right) & = & \left(%
\begin{array}{ccc}
  0 & m_D & 0 \\
  m_D & 0 & M_1 \\
  0 & M_1 & M_2 \\
\end{array}%
\right)
\left(%
\begin{array}{c}
  \nu \\
  \nu^c \\
  N \\
\end{array}%
\right). \eea  Note, $|{\rm Det} {\cal M}| =  m_D^2 M_2$, Tr
${\cal M} = M_2$ and $m_D \ll M_1, M_2$.  Hence the effective
See-Saw scale $M_{\rm eff}$ in this case may be obtained by
evaluating the inverse of the heavy $2 \times 2$ mass matrix  \bea
\left(
\begin{array}{cc}
   0 & M_1 \\
   M_1 & M_2 \\
\end{array} \right) \eea
in the $\nu^c$ direction.  We find $M_{\rm eff} = M_1^2/M_2$.
Note, the result is independent of the mass ordering, i.e. $M_1
\ll M_2$, $M_1 \gg M_2$ or $M_1 \approx M_2$.  Finally, we obtain
the light neutrino mass given by \bea m_{\nu} & \simeq & \f{m_D^2
M_2}{M_1^2}, \eea irrespective of the ratio $M_1/M_2$, as long as
$m_D \ll M_1, M_2$.

Note that the effective See-Saw mass $M_{\rm eff}$, in either
case, can be lower than the cutoff scale of the theory. For
example, in 4D, the natural size of the right-handed tau neutrino
Majorana mass is determined by taking the cutoff scale $M_* =
M_{Pl}$ in the effective higher dimension operator. By replacing
$\la \overline{16} \ra \sim M_{GUT}$ one obtains $M_{\nu^c} \equiv
M_{\rm eff} \sim \f{M_{GUT}^2}{M_{Pl}} \sim 10^{14} \ \GeV$.

In 5D, we have several possible choices for locating the matter
multiplets on the SO(10) or PS brane or in the bulk.  We also can
imagine SO(10) singlet fields, living in the bulk and giving light
neutrino masses via a double See-Saw mechanism.  We have
considered a detailed analysis of all possible ways of obtaining
neutrino masses in 5D SO(10) models~\cite{Kim:2003vr}.  The bottom
line can be obtained with simple dimensional analysis.

Consider an effective higher dimensional right-handed neutrino
mass operator on the PS brane given by \bea W & = & \f{C}{2
M_{*}^n} \bf \bar \chi^c \ \psi^c \ \bar \chi^c \ \psi^c \;
\delta(y - \f{\pi R}{2}). \eea  The constant $C$ is fixed by Naive
Dimensional Analysis with $C = 16 \pi^2 c$ and $n$ = 1  (for
$\psi^c$ on the PS brane) or $C = 24 \pi^3 c$ and $n$ = 2  (for
$\psi^c$ in the bulk). Thus when $4 \pi \la \bar \chi^c \ra \sim
M_*$  in the right-handed neutrino direction and $c \sim 1$ (i.e.
the ``natural" values) we find a right-handed neutrino mass \bea W
& = & \f{1}{2} M_{\rm eff} \ \nu^c \ \nu^c \eea  with \bea M_{\rm
eff} & \sim & \left(
\begin{array}{c} M_* \\ \f{3}{4} M_c \end{array} \right) \;\; {\rm for} \;\; \psi^c
\;\; \left( \begin{array}{c}    {\rm on \; the \; PS \; brane}  \\
\;\; {\rm in \; the \; bulk} \end{array} \right). \eea

We find that the value of $M_{\rm eff}$ for $\psi^c$ in the bulk
is of order $M_c$, but for $\psi^c$ on the PS brane we can only
obtain this desired value with a small value of $c \sim 10^{-3}$.
Thus for $\psi^c$ on the PS brane we need to suppress $M_{eff}$.
This can be accomplished in two possible ways.  In the first way,
by assuming a spontaneously broken U(1) symmetry which suppresses
the effective operator coefficient $C$.  This suppression can be
the consequence of a U(1) symmetry requiring the insertion of a
singlet field $S$ in the effective operator. Then a suppression
factor $\la S \ra/M_*$ is obtained. In the second case, a Majorana
mass $ \ll M_c$ is assumed for the bulk singlet field. This seems
to be the only situation where the Majorana mass of an SO(10)
singlet, located in the bulk, is important.

\subsection{CONCLUSION --- Theoretical Score Card}

Let me conclude by listing the virtues and problems of 5D SUSY
SO(10).  \bigskip

{\bf Virtues of the 5D theory} \medskip
\begin{itemize}
\item Charge quantization \& Family structure --- $\surd$
\medskip

\item Gauge coupling unification --- $\surd$
\medskip

\item Yukawa coupling unification for the third generation ---
$\surd$
\medskip

\item R parity $\Longrightarrow$ dark matter candidate --- $\surd$
\medskip

\item Neutrino mass (See--Saw mechanism) ---  $\surd$
\medskip

\item Gauge symmetry breaking --- $\surd$
\medskip

\item Higgs doublet--triplet splitting --- $\surd$
\medskip

\item Proton decay ($p \rightarrow K^+ \ \bar \nu$) due to dim. 5
Operators --- {\it R symmetry prevents dim. 5 ops.}
--- $\surd$
\end{itemize}

\bigskip \noindent {\bf Problems of the 5D theory}
\medskip
\begin{itemize}
\item Right-handed neutrino mass scale of order $ M_c$ can be
obtained.  It is only natural if the right-handed components of
quarks and leptons are located in the bulk.
\medskip

\item Proton decay ($p \rightarrow e^+ \ \pi^0$) due to dim. 6
operators --- {\it negligible in 4D, however in 5D one is now
sensitive to physics at the cutoff and the effects are
incalculable (and perhaps even observable ?)}

\end{itemize}


\begin{thebibliography}{99}

\bibitem{Kim:2002im}
H.~D.~Kim and S.~Raby,
JHEP {\bf 0301}, 056 (2003) [arXiv:hep-ph/0212348].

\bibitem{Kim:2003vr}
H.~D.~Kim and S.~Raby,
JHEP {\bf 0307}, 014 (2003) [arXiv:hep-ph/0304104].

\bibitem{Dermisek:2001hp}
R.~Dermisek and A.~Mafi,
Phys.\ Rev.\ D {\bf 65}, 055002 (2002) [arXiv:hep-ph/0108139].

\bibitem{Asaka:2001eh}
T.~Asaka, W.~Buchmuller and L.~Covi,
Phys.\ Lett.\ B {\bf 523}, 199 (2001) [arXiv:hep-ph/0108021];

L.~J.~Hall, Y.~Nomura, T.~Okui and D.~R.~Smith,
Phys.\ Rev.\ D {\bf 65}, 035008 (2002) [arXiv:hep-ph/0108071];

T.~j.~Li,
Nucl.\ Phys.\ B {\bf 619}, 75 (2001) [arXiv:hep-ph/0108120];

L.~J.~Hall and Y.~Nomura,
arXiv:hep-ph/0207079.

\bibitem{Nomura:2001mf}
Y.~Nomura, D.~R.~Smith and N.~Weiner,
Nucl.\ Phys.\ B {\bf 613}, 147 (2001) [arXiv:hep-ph/0104041].

\bibitem{Hall:2001pg}
L.~J.~Hall and Y.~Nomura,
Phys.\ Rev.\ D {\bf 64}, 055003 (2001) [arXiv:hep-ph/0103125];

R.~Contino, L.~Pilo, R.~Rattazzi and E.~Trincherini,
Nucl.\ Phys.\ B {\bf 622}, 227 (2002) [arXiv:hep-ph/0108102].

L.~J.~Hall and Y.~Nomura,
Phys.\ Rev.\ D {\bf 65}, 125012 (2002) [arXiv:hep-ph/0111068].

\bibitem{Murayama:2001ur}
H.~Murayama and A.~Pierce,
Phys.\ Rev.\ D {\bf 65}, 055009 (2002) [arXiv:hep-ph/0108104].

\bibitem{doubleseesaw}  R.N. Mohapatra and J.W.F. Valle, {\it Phys. Rev.}{\bf D34}
1634 (1986).

\end{thebibliography}
\end{document}